\documentclass{Interspeech}

\interspeechcameraready

\title{A Study on Speech Assessment with Visual Cues}

\author[affiliation={1,2}]{Shafique}{Ahmed}
\author[affiliation={1}]{Ryandhimas}{E. Zezario}
\author[affiliation={3}]{Nasir}{Saleem}
\author[affiliation={3}]{Amir}{Hussain} 
\author[affiliation={4}]{Hsin-Min}{Wang}
\author[affiliation={1}]{\\Yu}{Tsao}

\affiliation{Research Center for Information Technology Innovation}{Academia Sinica}{Taiwan}
\affiliation{}{National Tsing Hua University}{Taiwan}
\affiliation{}{Edinburgh Napier University, Scotland}{UK}
\affiliation{Institute of Information Science}{Academia Sinica}{Taiwan}

\email{shafique.khattak13@citi.sinica.edu.tw, yu.tsao@citi.sinica.edu.tw}

\keywords{speech assessment, multimodal learning, non-intrusive speech assessment, speech quality estimation}

\usepackage{comment}
\usepackage{multirow}
\begin{document}

\maketitle





\begin{abstract}
Non-intrusive assessment of speech quality and intelligibility is essential when clean reference signals are unavailable. In this work, we propose a multimodal framework that integrates audio features and visual cues to predict PESQ and STOI scores. It employs a dual-branch architecture, where spectral features are extracted using STFT, and visual embeddings are obtained via a visual encoder. These features are then fused and processed by a CNN-BLSTM with attention, followed by multi-task learning to simultaneously predict PESQ and STOI. Evaluations on the LRS3-TED dataset, augmented with noise from the DEMAND corpus, show that our model outperforms the audio-only baseline. Under seen noise conditions, it improves LCC by 9.61\% (0.8397→0.9205) for PESQ and 11.47\% (0.7403→0.8253) for STOI. These results highlight the effectiveness of incorporating visual cues in enhancing the accuracy of non-intrusive speech assessment.
\end{abstract}

\section{Introduction}

Objective assessment of speech quality and intelligibility is crucial for various applications, ranging from optimizing telecommunications systems \cite{Beerends13-POLQA} evaluating speech enhancement algorithms \cite{Chao24-Mamba, Defossez20-RTSE} and to developing effective assistive technologies such as hearing aids \cite{Diehl23-HearingAids, KirtonWingate23-SNR}. Traditional intrusive assessment methods, such as the Perceptual Evaluation of Speech Quality (PESQ) \cite{Rix01-PESQ} and Short-Time Objective Intelligibility (STOI) \cite{Taal11-STOI}, rely on direct comparisons between degraded speech signals and clean references. However, clean reference signals are rarely available in real-world scenarios, requiring the development of non-intrusive techniques that can predict speech quality and intelligibility directly from the degraded signal.

Recent advances in deep learning have shown notable progress in non-intrusive speech quality assessment \cite{Zezario23-MOSANet, dnsmos, urgent, cnn_ni, NISQA, speechbert, mosbench}. By training deep learning models on large-scale datasets of degraded speech paired with subjective or objective quality scores as ground truth, these models can capture diverse acoustic information, resulting in higher prediction performance. However, their performance often drops in unseen environments, particularly under low signal-to-noise ratio (SNR) conditions. For instance, our experimental results have shown that the correlation coefficient score can drop by up to 13\% between seen and unseen scenarios. This limitation is critical because many real-world applications (e.g., mobile communications in crowded areas or hearing aids in noisy cafes) inherently operate in low-SNR conditions, where inaccurate quality assessment may cause device misconfigurations and reduced overall intelligibility.

Motivated by the success of multimodal learning in related speech tasks, we intend to explore its application for speech quality assessment. In speech enhancement \cite{Ephrat18-LookListen, Ahmed24-DCUNet} and automatic speech recognition (ASR) \cite{Afouras18-AVSR}, integrating auxiliary modalities—particularly visual cues—has proven highly useful. Lip movements, for example, provide complementary and noise-robust information that can compensate for degraded acoustic signals by conveying clear articulatory and phonetic cues \cite{tseng2020study}. Although visual information has been successfully explored for tasks like noise suppression and speech separation \cite{Ephrat18-LookListen}, its potential for non-intrusive speech quality assessment remains largely unexplored.

In this work, we introduce a novel multimodal non-intrusive speech assessment model that leverages visual cues, specifically lip movements, to enhance the prediction of objective speech quality and intelligibility metrics, namely PESQ and STOI. Building on established CNN-BLSTM architectures with attention mechanisms \cite{Zezario20-STOINet, Zezario23-MOSANet}, which effectively capture temporal dependencies and audio features, our approach extends these models by incorporating visual embeddings extracted from video frames. We employ a convolutional recurrent network (CRN) \cite{Tan18-CRN} with attention mechanisms \cite{attn} to selectively focus on the important audio and visual features at each time step. We hypothesize that adding the visual modality provides important complementary information, especially in low-SNR conditions where relying solely on audio features becomes unreliable.


To evaluate our proposed approach, we first compare audio-only and multimodal models for PESQ and STOI prediction using a single-task learning framework. Following this, we explore a multi-task learning approach that jointly predicts both PESQ and STOI metrics, aiming to further examine the robustness and generalization. Our experiments are conducted on the LRS3 dataset \cite{Afouras18-LRS3}, augmented with different noise types from the DEMAND database \cite{Thiemann13-DEMAND} to simulate realistic acoustic degradations. The model's performance is evaluated using the Linear Correlation Coefficient (LCC), Spearman’s Rank Correlation Coefficient (SRCC), and Mean Squared Error (MSE) under both 'seen' (noise types present in the training data) and 'unseen' (noise types not included in the training data) conditions. Experimental results confirm that incorporating visual embeddings significantly improves assessment prediction accuracy, particularly in challenging, unseen noise conditions. The LCC score increases from 0.6336 to 0.8040 when predicting PESQ and from 0.6435 to 0.7692 when predicting STOI. This further confirms the importance of multimodal integration for robust speech assessment.



The remainder of this paper is organized as follows. Section 2. presents the proposed methodology. Section 3. describes the experimental setup and results. Finally, Section 4. concludes the paper and outlines directions for future work.

\section{Methodology}

In this work, we propose a multimodal non-intrusive speech assessment model that combines the acoustic and visual cues to robustly predict objective speech quality and intelligibility scores. Given a speech waveform $(X = [x_1, x_2, \ldots, x_T])$ and corresponding visual cues video $(V = [v_1, v_2, \ldots,v_M]$), the model processes the inputs through two parallel branches. In the first branch, the speech signal is transformed by the short-time Fourier transform (STFT) to extract its magnitude spectrogram 

\begin{equation}
S = \left|\mathrm{STFT}(X)\right| \in \mathbb{R}^{T \times F},\\
\end{equation}
where \(T\) denotes the number of time frames and \(F\) the number of frequency bins.

In parallel, the visual encoder extracts information from lip movements. The pre-processed video frames \(V\) (e.g., resized and normalized)  are fed into the visual encoder, which comprises a 3D convolutional layer followed by a ResNet-18 network, extracting deep visual embeddings, as detailed below:

\begin{equation}
F_v = \operatorname{ResNet\text{-}18}(\operatorname{Conv3D}(V)) \in \mathbb{R}^{M \times d_v}.\\
\end{equation}
Where, \( M \) denotes the number of video frames, \( d_v \) represents the dimensionality of the visual feature embeddings.
To synchronize the modalities, the visual features are up-sampled (e.g., using linear interpolation) to match the \(T\) frames of the audio spectrogram, resulting in 
\begin{equation}
\tilde{F}_v \in \mathbb{R}^{T \times d_v}.\\
\end{equation}
The multimodal fusion is then achieved by concatenating the audio and up-sampled visual features along the feature dimension:
\begin{equation}
F = \operatorname{concat}(S, \tilde{F}_v) \in \mathbb{R}^{T \times (F + d_v)}.\\
\end{equation}

\begin{figure*}[t]
  \centering
  \includegraphics[width=17cm]{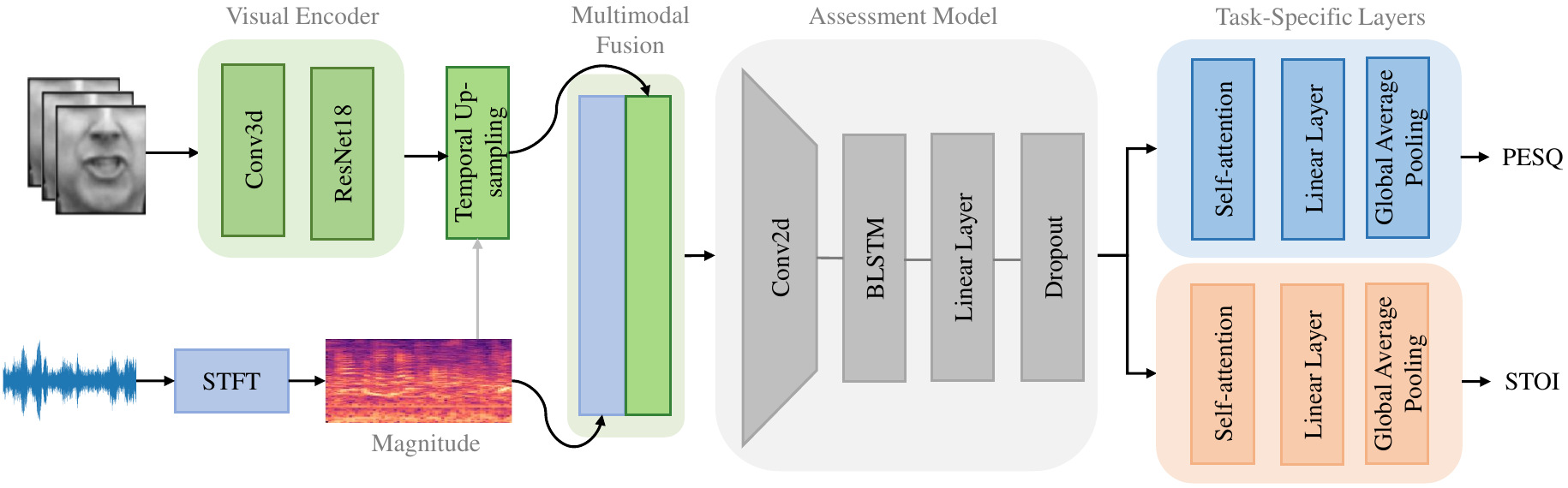}
  \caption{Overview of the Proposed Model Architecture}
  \label{fig:model_architecture}
\end{figure*}

Figure~\ref{fig:model_architecture} illustrates the overall architecture of the model. The fused representation \(F\) is forwarded to the speech assessment block, which is built on a convolutional recurrent neural network (CRNN). This block consists of multiple convolutional layers, followed by bidirectional LSTM layers, dense layer, and dropout to extract the assessment block features
\begin{equation}
H = \operatorname{CRNN}(F) \in \mathbb{R}^{T \times d_h},
\end{equation}
that capture both local spectral-temporal details and long-term dependencies. Next, each frame-wise latent feature \(H\) is processed by a task-specific branch, which first applies a self-attention layer to specifically focus on the more important region of the features. The resulting attention feature \(\tilde{H}\) is then passed through a linear layer to obtain a frame-level prediction $\tilde{Q}=[\tilde{q}_1, \tilde{q}_2, \ldots, \tilde{q}_T]$. The frame-level predictions are then aggregated through a global average pooling operation, producing a single utterance-level score:
\begin{equation}
\hat{Q} = \frac{1}{T} \sum_{t=1}^{T} \tilde{q}_t.
\end{equation}
Note that in the single-task learning framework, a dedicated task-specific branch predicts either the PESQ or STOI score. In contrast, the multi-task framework employs two separate branches to predict both scores simultaneously.

Furthermore, to effectively train the model with utterance-level prediction, we adopt a loss function that combines utterance-level and frame-level loss \cite{mosnet}. Let \(Q_n\) and \(I_n\) denote the ground truth utterance-level scores for PESQ and STOI, respectively, and let \(\hat{Q}_n\) and \(\hat{I}_n\) denote the corresponding predicted scores. Similarly, let \(\hat{q}_{n,l}\) and \(\hat{i}_{n,l}\) represent the frame-level predictions for the \(l\)-th frame of the \(n\)-th utterance, and let \(L(U_n)\) be the number of frames in the \(n\)th utterance. The loss for PESQ prediction is defined as


\begin{equation}
\resizebox{\linewidth}{!}{$
\mathcal{L}_{\text{PESQ}} = \frac{1}{N} \sum_{n=1}^{N} \left[ (Q_n - \hat{Q}_n)^2 + \alpha_Q \frac{1}{L(U_n)} \sum_{l=1}^{L(U_n)} (Q_n - \hat{q}_{n,l})^2 \right]
$}
\end{equation}

and the loss for STOI prediction is defined as


\begin{equation}
\resizebox{\linewidth}{!}{$
\mathcal{L}_{\text{STOI}} = \frac{1}{N} \sum_{n=1}^{N} \left[ (I_n - \hat{I}_n)^2 + \alpha_I \frac{1}{L(U_n)} \sum_{l=1}^{L(U_n)} (I_n - \hat{i}_{n,l})^2 \right]
$}
\end{equation}

where \(\alpha_Q\) and \(\alpha_I\) are hyperparameters that balance the contributions of the frame-level and utterance-level losses. In the multi-task configuration, the overall loss is computed as a weighted sum:
\begin{equation}
\mathcal{L}_{\text{Total}} = \beta \mathcal{L}_{\text{PESQ}} + \gamma \mathcal{L}_{\text{STOI}},
\end{equation}

with \(\beta\) and \(\gamma\) determine the relative importance of each task.


\section{Experiments}

\subsection{Experimental Setup}
We evaluate our proposed multimodal non-intrusive speech assessment model using the LRS3-TED dataset, a large-scale audiovisual corpus containing thousands of utterances extracted from TED and TEDx videos. For our experiments, we randomly selected one utterance per speaker folder, resulting in 4004 clean training utterances and 412 clean test utterances.

To simulate realistic acoustic degradations, clean speech is augmented with noise from the DEMAND corpus. Noise samples are divided into seen and unseen categories using 80–20 ratio. For the training set, noisy utterances are generated by mixing clean speech with seen noise types from the DEMAND corpus at signal-to-noise ratios (SNR) ranging from –20 dB to 10 dB, resulting in 40,400 noisy utterances. These noisy utterances are split into two equal parts. One part is directly included in the training set as noisy speech, while the other is processed using different deep learning-based speech enhancement models to emulate enhanced speech conditions. Specifically, the second set of noisy data is divided into three equal portions, each enhanced with a different model selected for its unique feature domain. First, the Denoiser model from Facebook Research \cite{Defossez20-RTSE} processes the raw waveform. Next, the SE-Mamba model \cite{Chao24-Mamba} leverages phase information in the complex domain. Finally, a simple LSTM-based architecture is applied to enhance spectral features. Consequently, the training set comprises an additional 20,200 enhanced utterances, with each enhancement method applied uniformly to a subset of the noisy data. The test set is constructed similarly, consisting of 4120 seen noisy utterances (and their corresponding enhanced versions), as well as 1648 unseen noisy utterances (with their enhanced utterances).

For audio feature extraction, we compute the STFT of the raw speech signal using a window length of 512, an FFT size of 512 and a window increment of 256 samples. In parallel, visual feature extraction involves extracting the lip region from each video frame, resizing it to 88×88 pixels to highlight visual cues associated with lip movements. In addition, the frames were converted to grayscale to reduce computational complexity.

For the model configuration, the visual encoder includes a 3D convolutional layer with 64 channels followed by a ResNet-18 backbone to extract visual embeddings. The speech assessment model comprises 12 convolutional layers with channel configurations of \(\{16, 32, 64, 128\}\), followed by a bidirectional LSTM layer with 128 nodes. This is connected to a linear layer with 128 units and a dropout layer with a rate of 0.3. The refined features are then passed to the task-specific layer, where a self-attention mechanism is applied to integrate contextual information and further refine feature representations. Finally, a linear layer with one neuron produces a frame-wise prediction, and a global average pooling operation aggregates these predictions to compute the final assessment score.

All models were trained using the same configuration to ensure consistency between experiments. These models were train for 25 epochs with an early stopping function to prevent overfitting. For parameter optimization, we used Adam optimizer with an initial learning rate of \(1 \times 10^{-4}\) and used a learning rate scheduler that reduced the learning rate by a factor of 0.1 when validation performance plateaued.

\subsection{Evaluation Results}
Our model's performance is evaluated using three commonly used standard metrics \cite{mosnet, Zezario23-MOSANet}: linear correlation coefficient (LCC), spearman’s rank correlation coefficient (SRCC), and mean squared error (MSE). In our evaluation, a lower MSE indicates that the predicted scores are closer to the ground truth (lower, the better), whereas higher LCC and SRCC values reflect a stronger correlation between predicted and ground truth scores (higher, the better). We evaluated the models under both seen and unseen noise conditions.

Table~\ref{tab:performance_single_task} summarizes the performance of our single-task learning framework, where separate models are trained to predict the PESQ and STOI scores. The multimodal architecture shows significant improvements over the audio-only models for each task. For instance, under seen noise conditions, the multimodal model improves the PESQ LCC from 0.8316 to 0.9210 and the STOI LCC from 0.7662 to 0.8436. Under unseen noise conditions, the PESQ LCC increases from 0.6177 to 0.8056 and the STOI LCC from 0.6921 to 0.7925, while MSE values are considerably reduced, showing the improved prediction performance.

\begin{table}[!t]
  \centering
  \caption{Detailed evaluation metrics (PESQ and STOI) for single-task learning under seen and unseen noise conditions.}
  \label{tab:performance_single_task}
  \scriptsize
  \setlength{\tabcolsep}{3.0pt}
  \resizebox{\columnwidth}{!}{%
    \begin{tabular}{llcccccc}
      \toprule
      \multirow{2}{*}{Condition} & \multirow{2}{*}{Modality} & \multicolumn{3}{c}{PESQ} & \multicolumn{3}{c}{STOI} \\
      \cmidrule(lr){3-5}\cmidrule(lr){6-8}
      & & LCC \(\uparrow\) & SRCC \(\uparrow\) & MSE \(\downarrow\) & LCC \(\uparrow\) & SRCC \(\uparrow\) & MSE \(\downarrow\) \\
      \midrule
      \multirow{2}{*}{Seen Noise} 
        & Audio-Only   & 0.8316 & 0.8155 & 0.3855  & 0.7662 & 0.7660 & 0.0053 \\
        & Multimodal   & \textbf{0.9210} & \textbf{0.9086} & \textbf{0.1182} & \textbf{0.8436} & \textbf{0.8376} & \textbf{0.0038} \\
      \multirow{2}{*}{Unseen Noise}
        & Audio-Only   & 0.6177 & 0.6875 & 0.2079 & 0.6921 & 0.6839 & 0.0107 \\
        & Multimodal   & \textbf{0.8056} & \textbf{0.8323} & \textbf{0.1191} & \textbf{0.7925} & \textbf{0.7867} & \textbf{0.0078} \\
      \bottomrule
    \end{tabular}%
  }
\end{table}

Table~\ref{tab:performance_multi_task} provides a detailed evaluation of the multi-task learning approach, where a single model is jointly trained for multiple tasks—specifically, predicting both PESQ and STOI. Similar to the single-task scenario, the multi-task approach with multimodal inputs shows improvements over the single-modality (audio-only) model. By looking at the results, we observe that for the test set containing unseen noisy and enhanced utterances, the multimodal approach improves the LCC for PESQ from 0.6336 to 0.8040 and for STOI from 0.6435 to 0.7692. Under seen noise conditions, the LCC for PESQ increases from 0.8397 to 0.9205, and for STOI from 0.7403 to 0.8253. Overall, these results show the effectiveness of the multimodal approach.

\begin{table}[!t]
  \centering
  \caption{Detailed evaluation metrics (PESQ and STOI) for Multi-task learning under seen and unseen noise conditions.}
  \label{tab:performance_multi_task}
  \scriptsize
  \setlength{\tabcolsep}{3.0pt}
  \resizebox{\columnwidth}{!}{%
    \begin{tabular}{llcccccc}
      \toprule
      \multirow{2}{*}{Condition} & \multirow{2}{*}{Modality} & \multicolumn{3}{c}{PESQ} & \multicolumn{3}{c}{STOI} \\
      \cmidrule(lr){3-5}\cmidrule(lr){6-8}
      & & LCC \(\uparrow\) & SRCC \(\uparrow\) & MSE \(\downarrow\) & LCC \(\uparrow\) & SRCC \(\uparrow\) & MSE \(\downarrow\) \\
      \midrule
      \multirow{2}{*}{Seen Noise} 
        & Audio-Only   & 0.8397 & 0.8288 & 0.3205 & 0.7403 & 0.7552 & 0.0060 \\
        & Multimodal   & \textbf{0.9205} & \textbf{0.9078} & \textbf{0.1221} & \textbf{0.8253} & \textbf{0.8487} & \textbf{0.0041} \\
      \multirow{2}{*}{Unseen Noise}
        & Audio-Only   & 0.6336 & 0.7042 & 0.1873 & 0.6435 & 0.6530 & 0.0112 \\
        & Multimodal   & \textbf{0.8040} & \textbf{0.8400} & \textbf{0.1152} & \textbf{0.7692} & \textbf{0.7813} & \textbf{0.0081} \\
      \bottomrule
    \end{tabular}%
  }
\end{table}

\begin{figure}[t]
  \centering
  \includegraphics[width=\linewidth]{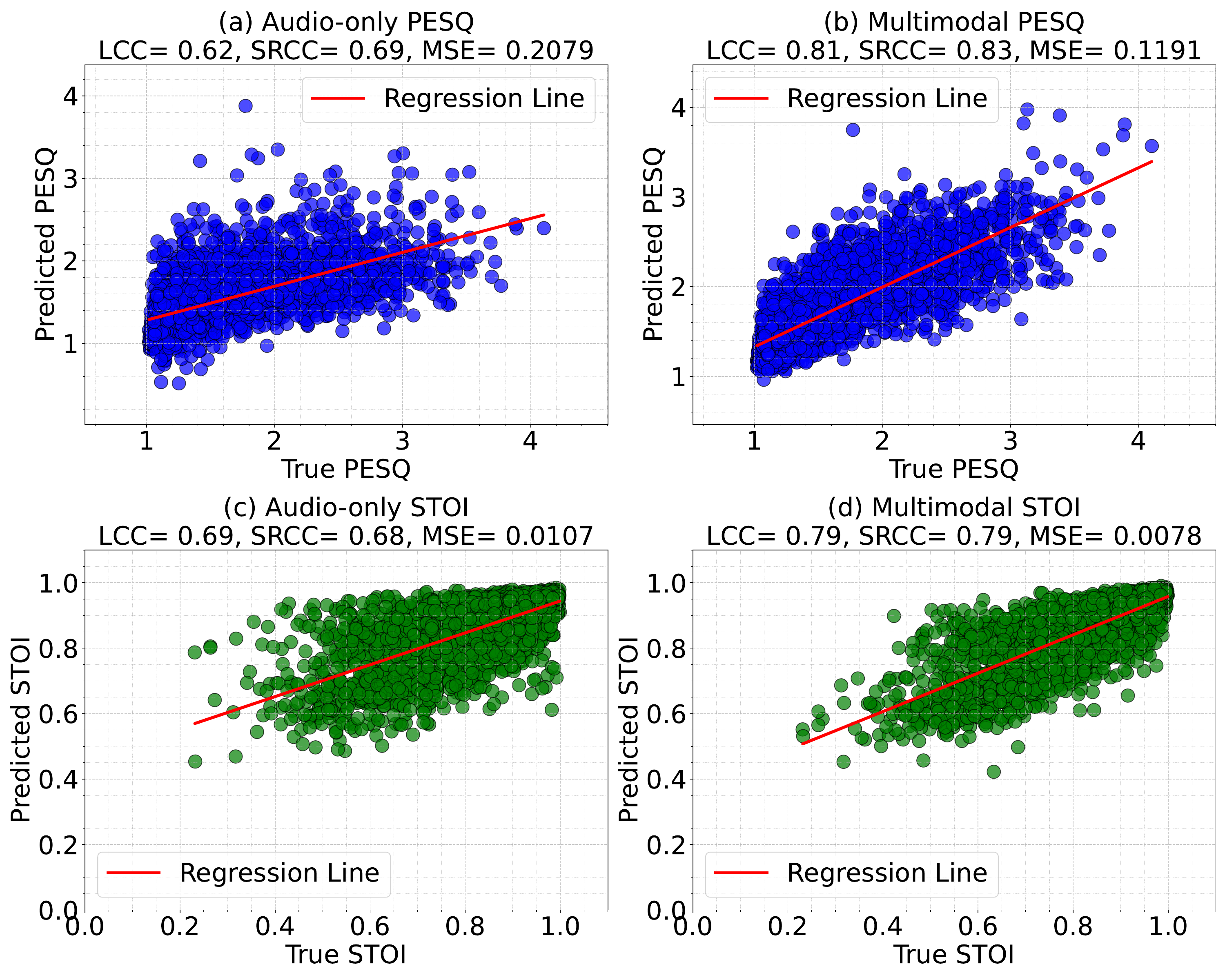}
  \caption{Scatter Plots for Single-Task Learning under Unseen Noise Conditions. (a) Audio-only PESQ, (b) Multimodal PESQ, (c)Audio-only STOI, and (d) Multimodal STOI}
  \label{fig:scatter_plot_single-task}
\end{figure}
\begin{figure}[t]
  \centering
  \includegraphics[width=\linewidth]{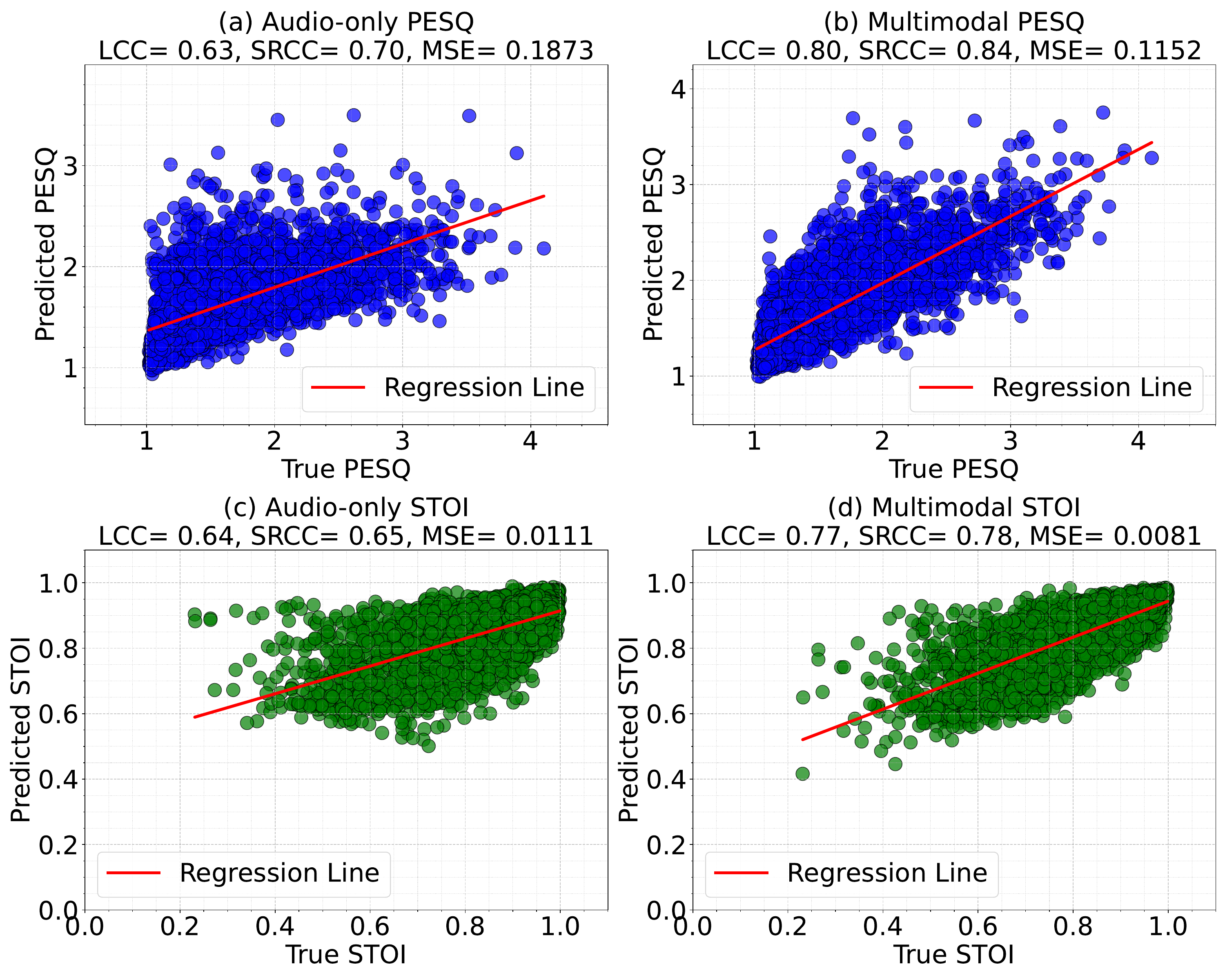}
  \caption{Scatter Plot for Multi-Task Learning under Unseen Noise Conditions. (a)Audio-only PESQ, (b) Multimodal PESQ, (c) Audio-only STOI, and (d) Multimodal STOI.}
  \label{fig:scatter_plot_multi-task}
\end{figure}

To visually represent our results, we have included a scatter plot that shows the correlation between the predicted assessment scores and the ground truth scores. Figure~\ref{fig:scatter_plot_single-task} shows the scatter plots for the single-task learning scenario under unseen noise conditions. Figure (a) displays the Audio-only PESQ predictions, (b) shows the Multimodal PESQ predictions, (c) shows the Audio-only STOI predictions, and (d) shows the Multimodal STOI predictions. Similarly, Figure~\ref{fig:scatter_plot_multi-task} shows the corresponding scatter plots for the multi-task learning framework with the same sub-figure layout. In these scatter plots, each point represents an individual test utterance with its predicted score (y-axis) plotted against the actual, ground-truth score (x-axis). In the multimodal subplots, the points cluster tightly along the diagonal, indicating that the prediction of the model closely matches the actual scores, indicating a strong correlation. In contrast, the audio-only subplots display a wider spreading of points around the diagonal, suggesting that the predictions are less consistent with the ground truth, which leads to weak correlation. These visualizations highlight the benefit of integrating visual cues into the speech assessment model.

\begin{figure}[t]
  \centering
  \includegraphics[width=\linewidth]{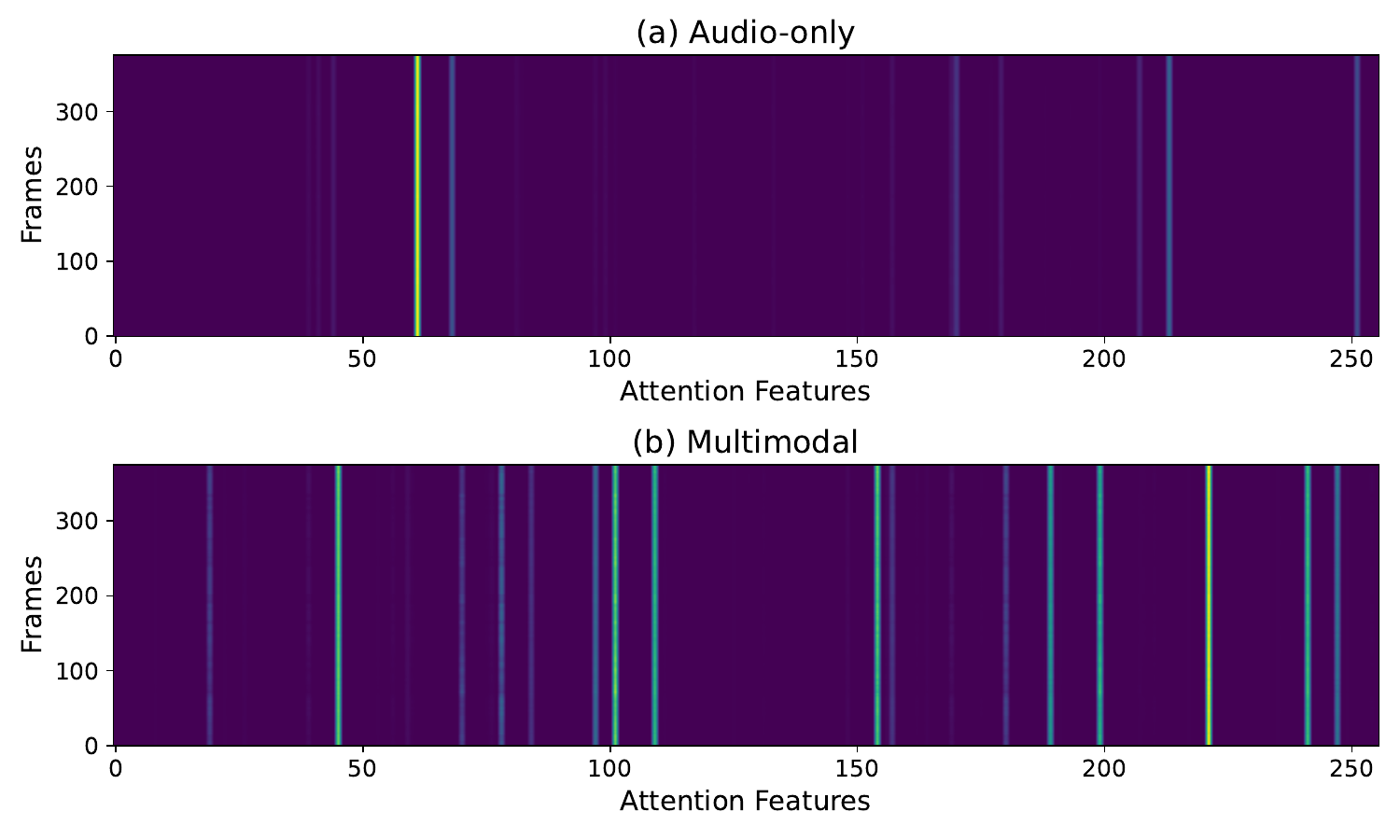}
  \caption{Latent representations of the attention layer: (a) Audio-only (b) Multimodal}
  \label{fig:attention}
\end{figure}

In addition to scatter plot analysis, we compared the attention layer feature representations of the audio-only and multimodal models. For visualization, we selected a single-task framework for STOI prediction to obtain both models' attention-layer latent representations. Figure \ref{fig:attention} presents the visual representation of the attention layer. As shown in Figure \ref{fig:attention} (a), the attention map for the audio-only model struggles to consistently focus on the most informative speech segments, while in Figure \ref{fig:attention} (b), the multimodal model exhibits a uniformly distributed attention pattern. This suggests that incorporating visual cues enables the model to learn shared, robust representations and assign more generalizable weights across the attention features.

Overall, these results confirm that leveraging visual cues with audio features improves the performance of non-intrusive speech assessment models. Integrating visual information provides additional context that enables the model to more accurately capture the aspects of speech assessment. Notably, multimodal models achieve lower MSE and higher LCC and SRCC values, especially in challenging low-SNR conditions and unseen noise scenarios.

\section{Conclusion}
In this paper, we introduced a novel multimodal non-intrusive speech assessment model that integrates acoustic and visual cues to robustly predict objective speech quality and intelligibility metrics such as PESQ and STOI. Extensive experiments on the LRS3-TED dataset, under realistic noisy and enhanced speech conditions, demonstrate that our multimodal approach significantly outperforms audio-only methods, especially in low-SNR and unseen noise scenarios. The single-task and multi-task learning configurations showed higher correlation coefficients and lower prediction errors, highlighting the benefit of using visual information for more reliable speech assessment.

For future work, we plan to extend our approach by incorporating self-supervised learning (SSL) models to further enhance feature extraction and representation learning. Recent studies have shown the effectiveness SSL in capturing robust speech representations, which have significantly improved speech assessment performance. In addition, our goal is to explore the broader potential of our framework by applying it to other multimodal speech processing tasks, such as audio-visual speech enhancement, audio-visual speech recognition, and integration into multimodal healthcare systems.

\bibliographystyle{IEEEtran}
\bibliography{mybib}

\end{document}